\documentclass[epj]{webofc}
\usepackage[varg]{txfonts}   
\woctitle{MESON2016 - the 14$^\textrm{th}$ International Workshop on Meson Production, Properties and Interaction}
\begin{document}
\selectlanguage{english}
\title{Constraints on atmospheric charmed-meson production from IceCube}

\author{Tomasz~Jan~Palczewski\inst{1,2}\fnsep\thanks{\email{tpalczewski@lbl.gov}}~for the IceCube Collaboration \footnote{\protect\url{http://icecube.wisc.edu}} 
}

\institute{University of California, Berkeley, USA
\and
               Lawrence Berkeley National Laboratory, Berkeley, USA          
          }

\abstract{

At very-high energies (100~TeV~-~1~PeV), the small value of Bjorken-x ($\le10^{-3}-10^{-7}$) at which the parton distribution functions are evaluated makes the calculation of charm quark production very difficult. The charm quark has mass ($\sim$1.5$\pm$0.2~GeV) significantly above the $\Lambda$$_{QCD}$ scale ($\sim$200~MeV), and therefore its production is perturbatively calculable. However, the uncertainty in the data and the calculations cannot exclude some smaller non-perturbative contribution. To evaluate the prompt neutrino flux, one needs to know the charm production cross-section in pN -> c$\bar{c}$ X, and hadronization of charm particles. This contribution briefly discusses computation of prompt neutrino flux and presents the strongest limit on prompt neutrino flux from IceCube. 
 
  
}
\maketitle
\section{Introduction}
\label{intro}

IceCube's detection of ultra-high energy neutrino events heralds the beginning of neutrino astronomy. 
The flux of cosmic-rays incident on the atmosphere has been measured by a variety of experiments~\cite{cite:cosmis-ray-exp1, cite:cosmis-ray-exp2, cite:cosmis-ray-exp3}. The dominant contribution to the neutrino flux at 10~GeV-1~TeV energies comes from the decays of charged pions and from leptonic and semi-leptonic decays of kaons~\cite{cite:pion-decays}. The neutrino flux from pions and kaons,   commonly referred to as the conventional neutrino flux, is well understood (steep energy spectrum $\sim$ E$^{-3.7}$ below $\sim$100~TeV). Prompt neutrinos, produced in the atmosphere by decays of charmed hadrons (e.g. D$^{0}$, D$^{\pm}$ $\rightarrow$ l $\nu_{e,\mu}$~X) that come from cosmic-ray interactions with atmospheric nuclei (pN $\rightarrow$ c$\bar{c}$~X), contribute to the atmospheric neutrino flux at high energies. The long-lived high energy pions and kaons interact before decaying into neutrinos. In contrast charmed particles have short lifetimes and decay into neutrinos independent of their energy up to $\sim$ 10$^{17}$~eV, and arrival direction. Therefore, the prompt neutrino energy spectrum should follow the spectrum of primary cosmic-rays (energy spectrum of $\sim$ E$^{-2.7}$ and an isotropic zenith distribution). The energy spectrum of astrophysical neutrinos is expected to follow the production spectrum at the cosmic-ray accelerator (in a simplistic approach: if Fermi shock acceleration is the responsible mechanism, a power law spectrum $\sim$E$^{-2}$ is expected).

\section{IceCube detector and detection principle}
\label{det}

IceCube is the world's largest neutrino detector, located near the geographic South Pole, instrumenting more than a cubic-kilometer of glacial ice. The detector is composed of Digital Optical Modules (DOMs) which are buried in ice, between depths of 1450m and 2450m, on 86 cables called "strings"~\cite{cite:detector}. On each string there are 60 DOMs. Each DOM consists of a 10-inch photomultiplier tube (PMT), calibration light sources, and digitizing electronics~\cite{cite:detector2}. 



      
\section{Estimation of prompt neutrino flux}
\label{estimation}

IceCube observations~\cite{cite:icecube1} indicate that the neutrino flux is dominated by conventional atmospheric neutrinos and by cosmic neutrinos at low energies and high energies, respectively. The precise knowledge of the prompt flux is of a key importance for the cosmic neutrino flux measurements. \par Many calculations of the prompt neutrino flux have been presented \cite{cite:charm-cal1, cite:charm-cal2, cite:charm-cal3, cite:charm-cal4, cite:charm-cal5, cite:charm-cal6}. The estimation of the prompt neutrino flux requires knowledge of the charm production cross-section ($\sigma_{c\bar{c}}$) in the process pN~$\rightarrow$ c$\bar{c}$~X and of the hadronization of charm particles. Theoretical predictions at high energies are uncertain due to uncertainties in charm mass, contributions from next-to-leading order (NLO) corrections, the factorization and renormalization scales, and the choice of parton distribution functions~\cite{cite:theory1, cite:theory2}. The charm quark is produced with high fraction of the momentum of the incoming cosmic-ray projectile. The cross-section is sensitive to the domain of parton densities at very small values of Bjorken-x ($\le10^{-3}-10^{-7}$). This region is poorly constrained by the experimental data. The small value of Bjorken-x at which the parton distribution functions are evaluated makes the calculation of charm production very difficult. The computation of the prompt neutrino flux also requires folding of the charmed hadronproduction cross-section with the flux of incoming cosmic-rays. This introduces uncertainty connected with the limited knowledge of the extremely high-energy cosmic ray composition. Recently, new updated parameterisations of the cosmic ray flux have been released~\cite{cite:cosmis-ray-exp1, cite:cosmic1, cite:cosmic2} with improved description above the 'knee' and more sophisticated composition description.
To obtain the prompt neutrino flux at the detector level the propagation of high energy particles and their decay products through the atmosphere needs to be described. The cascade formalism in the framework of the Z-momenta~\cite{cite:cascade-eq1} is commonly used for this evaluation. This approach requires, among others, as an input the total inelastic proton-Air cross-section which has large uncertainties at the high energy region~\cite{cite:total-inelastic-p-air}. In addition, the assumption that the charm production cross-section scales with the mean atomic number of air as compared to the corresponding pp cross-sections needs to be done. However, small contribution from the nuclear modification effects cannot be  fully rejected.  

\section{Validation of charm hadroproduction using LHC data}
\label{lhc_validation}
\begin{figure}[ht]
\centering
\sidecaption
\includegraphics[width=9.5cm,clip]{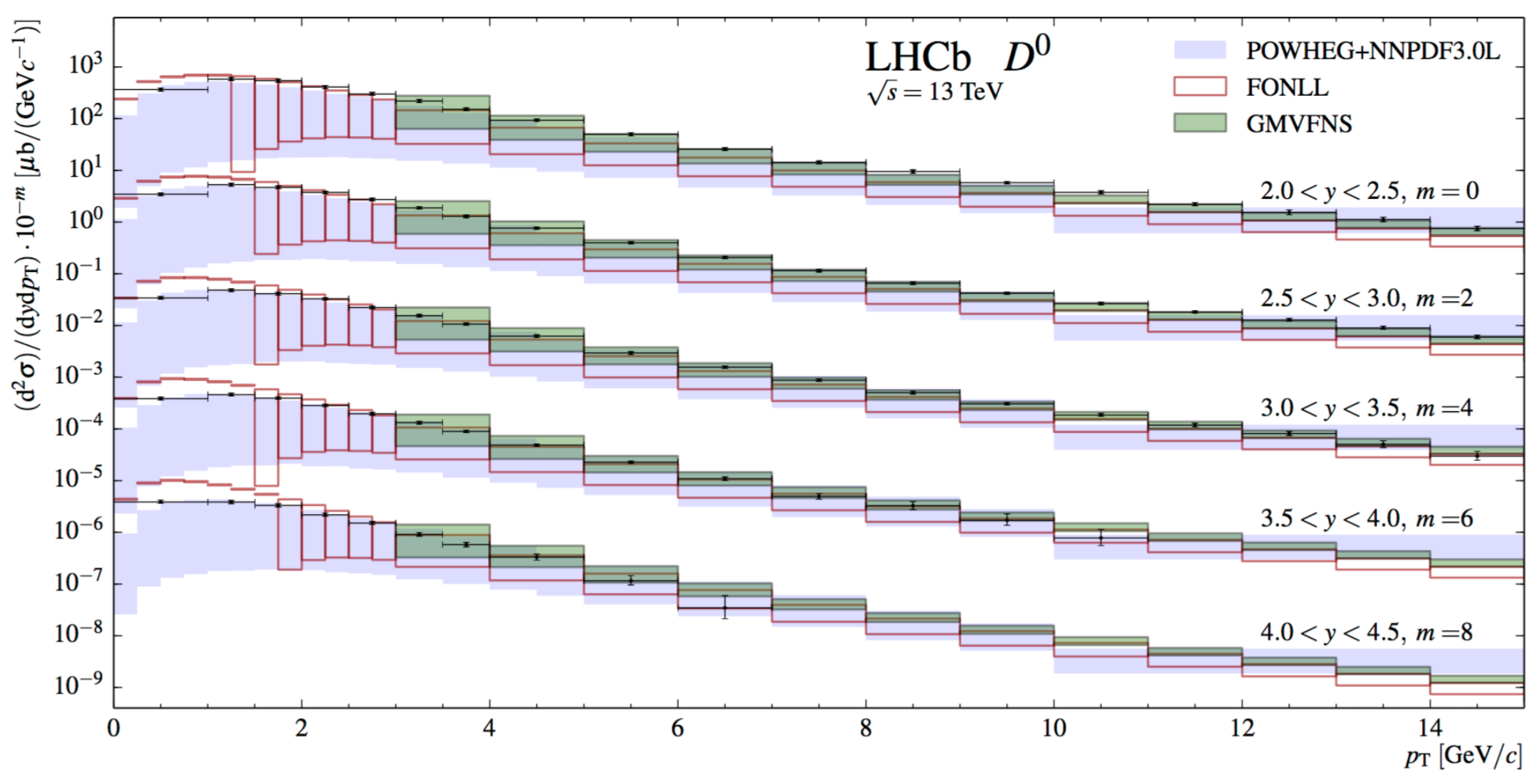}
\caption{Measurements and predictions for the absolute prompt D$^{0}$ cross-section at $\sqrt{s}$=13~TeV. The boxes indicate the $\pm$1$\sigma$ uncertainty band on the theory predictions. In cases where this band spans more than two orders of magnitude only its upper edge is indicated. Picture taken from~\cite{cite:lhc1}.}
\label{fig-3}       
\end{figure}

The LHC collider $\sqrt{s}$=7~TeV mode corresponds to a fixed target beam energy in pp collisions of 26 PeV. Therefore, the recent results from LHC~\cite{cite:lhc1, cite:lhc2, cite:lhc3, cite:lhc4} can constrain pQCD parameters (such as factorization and renormalization scales, parton distribution functions, estimates of cosmic ray spectra) which are crucial for the precise calculations of the charm hadroproduction. The charmed particle production cross-sections in pp collisions at a centre-of-mass energy of $\sqrt{s}$=7 and 13~TeV have been measured. As an example Fig.~\ref{fig-3} shows measurements and predictions for absolute prompt D$^{0}$ cross-sections from pp at $\sqrt{s}$=13~TeV (LHCb experiment). The shapes of differential cross-sections for D$^{0}$, D$^{+}$, D$^{*+}$, D$^{+}_{s}$ are in agreement with NLO predictions (the predicted central values are generally below the data but within the uncertainty). These precise LHC results are crucial to increase our knowledge in this field. Although, modern collider experiments are limited due to the lack of the coverage in the very large rapidity region. 

\section{Physics in the forward region}
\label{forward_region}

\begin{figure}[hbt]
\centering
\sidecaption
\includegraphics[width=7.cm,clip]{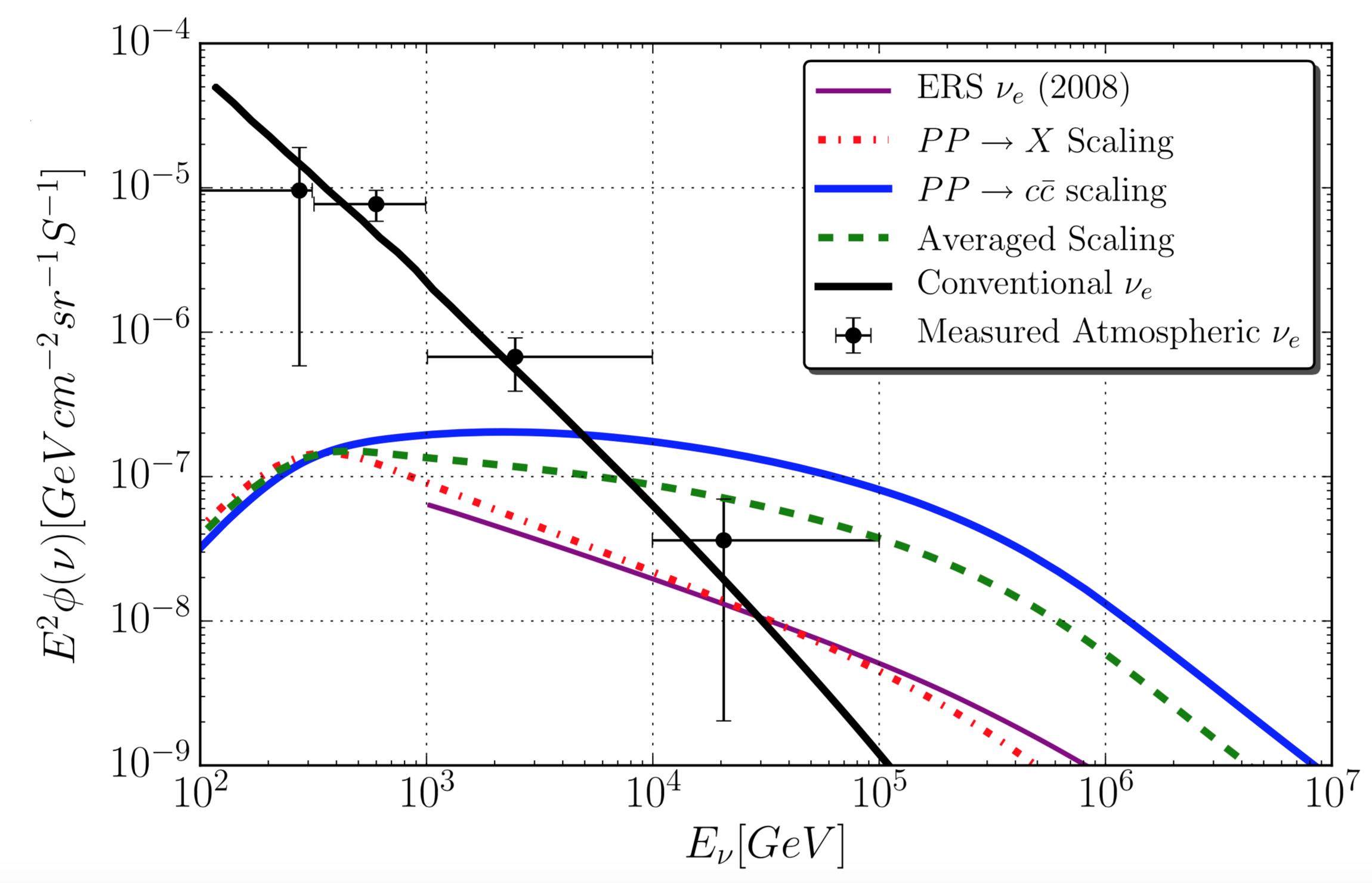}
\caption{The prompt electron neutrino spectrum from forward charm is shown for extreme assumptions of the energy dependence. Also shown is the result for an intermediate dependence that exceeds the measured flux~\cite{cite:33} at the 1$\sigma$ level at the highest energy of 20 TeV. An estimate of the contribution from centrally produced charm particles by~\cite{cite:33} is shown for comparison. The shown conventional atmospheric $\nu_{e}$ flux does not contain contribution from semileptonic decays of K$_{S}$ which becomes important, regardless of small branching ratios for these decays, at high energies~\cite{cite:spencer-gaisser}. For more details see~\cite{cite:halzen-logan}. Picture taken from~\cite{cite:halzen-logan}.}
\label{fig-4}       
\end{figure}

As it was pointed out in the Sect.~\ref{lhc_validation}, modern collider experiments, like experiments at RHIC and LHC, have limited coverage in the very large rapidity region. Consequently, the forward charmed particles are produced inside the beampipe. These particles dominate the high-energy atmospheric neutrino flux in underground experiments. In the literature, the forward charm production has been described in many ways and with different complexity, from intrinsic charm~\cite{cite:intrinsic-charm} to the inclusion of diffractive components in perturbative QCD~\cite{cite:diffractive-components}. In Ref.~\cite{cite:halzen-logan} authors showed that different forward physics assumptions modify the predictions of the prompt neutrino flux greatly (see Fig.~\ref{fig-4}).
   
\section{Prompt neutrino flux at IceCube }

\begin{figure}[ht]
\centering
\includegraphics[width=4.6cm,clip]{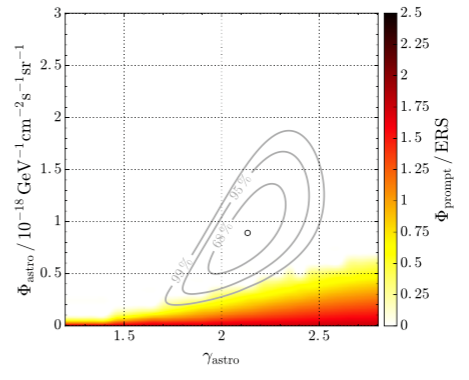}
\includegraphics[width=4.6cm,clip]{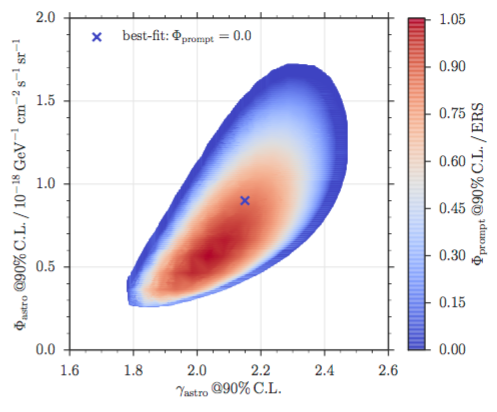}
\includegraphics[width=4.6cm,clip]{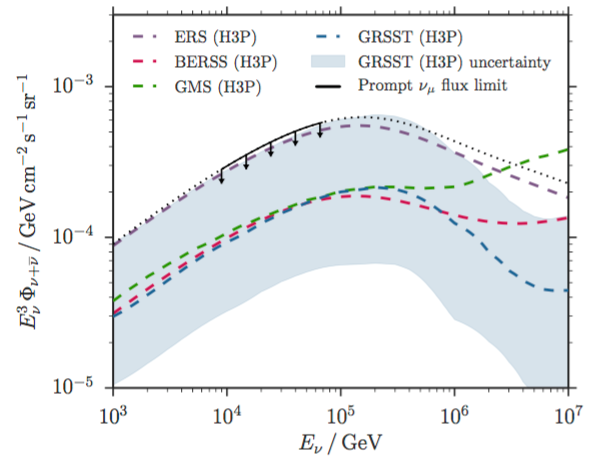}
\caption{Left: Best-fit prompt normalization in ERS units for each scan point $\Phi_{astro}$ and  $\gamma_{astro}$. Additionally, the two-dimensional contours for $\Phi_{astro}$ and  $\gamma_{astro}$ are shown; Middle: 90\% CL contour assuming Wilks? theorem based on a three dimensional profile likelihood scans of the astrophysical parameters $\Phi_{astro}$ and  $\gamma_{astro}$ and the prompt normalization in ERS units; Right: Prompt atmospheric muon neutrino flux predictions from selected  perturbative QCD calculations (dashed lines) in comparison to the upper limit from IceCube likelihood 6 year data analysis based on the up-going data sample. For more details see~\cite{cite:nu_mu-6y}. Pictures taken from~\cite{cite:nu_mu-6y}.}
\label{fig-prompt1}       
\end{figure}

One of the key IceCube physics goals is searching for astrophysical neutrinos and prompt atmospheric neutrinos in the measured data. The main background for this type of analyses consists of atmospheric muons and conventional atmospheric neutrinos. These backgrounds can be highly suppressed by selecting high-quality upward-going events. In the down-going event region the so called self-veto procedure~\cite{cite:self-veto} (an atmospheric neutrino is vetoed when accompanied by atmospheric muons) and high deposited charge cut can be used to suppress backgrounds as well as a selection of the starting events inside the detector (part of the detector is used as a veto). 
	Taking into account that neutrinos from charm have a harder spectrum than atmospheric neutrinos and that they are isotropic this provides an unique opportunity to confront the predicted prompt neutrino fluxes with the observed data. The recent binned likelihood analysis, with assumption of unbroken single power law astrophysical flux (described by two parameters: normalization $\Phi_{astro}$ at 100~TeV neutrino energy and the spectral index $\gamma_{astro}$) and per-bin expectations given by the sum of three components: conventional, prompt, and astrophysical, of up-going events in 6 years of data~\cite{cite:nu_mu-6y} gives to date the strongest prompt neutrino flux constraints by IceCube. The prompt normalization as a function of the normalization and spectral index is shown in Fig.~\ref{fig-prompt1}~(left). It is clearly seen that  non-zero prompt normalization appears only for strong deviations from the best fit astrophysical spectrum. Fig.~\ref{fig-prompt1}~(middle) shows the joint three-dimensional 90\% confidence region for the prompt flux and the astrophysical parameters. The maximum prompt flux in the three-dimensional confidence region is 1.06 $\times$ prediction taken from~\cite{cite:ers}~(ERS) and can be treated as conservative upper limit. Recent selected perturbative QCD calculations~\cite{cite:pqcd_m_2, cite:sarkar, cite:pqcd_m_1} are shown in Fig.~\ref{fig-prompt1}~(right) with the IceCube upper limit from the discussed likelihood 6 year data analysis. The GRSST (H3p) pQCD calculation is based on the same set of tools, POWHEG and NNPDF3.0L, used to calculate predictions of the absolute prompt particle cross-sections at LHC~\cite{cite:sarkar} (example shown in~Fig.~\ref{fig-3})            	
	
	IceCube has developed techniques for the extraction of physical measurements from atmospheric muon events that can be used to estimate the prompt component~\cite{cite:muon}. These techniques will not be discussed in this contribution. However, this is an important work as it gives independent verification  of the prompt measurement. Currently, a definite measurement of the prompt flux using muon data is not yet possible. Depending on which assumption is chosen for the systematic error, the final result varies considerably (for more details see~\cite{cite:muon}). In contrast to neutrinos, for muons additional decays of neutral vector mesons contribute to the prompt neutrino flux at an uncertain level~\cite{cite:muon-add, cite:muon-add2}. Once systematic effects are fully understood and controlled, it will be possible for the IceCube detector to measure the prompt contribution using muon data.     

\section{Summary}
\label{summary}

In the low-energy regime, atmospheric neutrino flux, which mainly comes from pion and kaon intermediate states, is well known. Proper calculation of the prompt neutrino flux is complex and depends on many theoretical aspects briefly discussed in this contribution. Atmospheric lepton flux is a background to cosmic neutrino flux searches but it is interesting in its own right as it probes low-x QCD. A precise measurement of prompt atmospheric muons and neutrinos could therefore complement the knowledge of differential cross sections for parton interactions and structure functions in accelerator physics. 

%

\begin{thebibliography}{00}
%
%


\bibitem{cite:cosmis-ray-exp1} T.~Stanev, T.~K.~Gaisser, S.~Tilav, Nucl. Instrum. Meth. A~\textbf{742},  42 (2014)

\bibitem{cite:cosmis-ray-exp2} E.~S.~Seo,  Astropart. Phys. \textbf{39}, 76 (2012)

\bibitem{cite:cosmis-ray-exp3} K.~H.~Kampert, M.~Unger, Astropart. Phys. \textbf{35}, 660 (2012)

\bibitem{cite:pion-decays} T.~K.~Gaisser, Earth Planets Space \textbf{62}, 195 (2010) 

\bibitem{cite:detector} R.~Abbasi~\textit{et al.}~(IceCube Collaboration), Nucl. Instrum. Meth. A~\textbf{601}, 294 (2009)

\bibitem{cite:detector2} M.~G.~Aartsen~\textit{et al.}~(IceCube Collaboration), JINST \textbf{9}, P03009 (2014)

\bibitem{cite:icecube1} M.~G.~Aartsen~\textit{et al.}~(IceCube Collaboration), Phys. Rev. D~\textbf{91}, 022001 (2015)

\bibitem{cite:charm-cal1} P.~Lipari, Astropart. Phys. \textbf{1}, 195 (1993)

\bibitem{cite:charm-cal2} L.~Pasquali, M.~Reno, I.~Sarcevic, Phys. Rev. D~\textbf{59}, 034020 (1999)

\bibitem{cite:charm-cal3} R.~Enberg, M.~H.~Reno, I.~Sarcevic, Phys. Rev. D\textbf{78}, 043005 (2008) 

\bibitem{cite:charm-cal4} P.~Gondolo, G.~Ingelman, M.~Thunman, Astropart. Phys. \textbf{5}, 309 (1996) 

\bibitem{cite:charm-cal5} A.~Martin, M.~Ryskin, A.~Stasto, Acta Phys. Polon. B~\textbf{34}, 3273 (2003) 

\bibitem{cite:charm-cal6} G.~Gelmini, P.~Gondolo, G.~Varieschi, Phys. Rev. D~\textbf{61}, 036005 (2000)

\bibitem{cite:theory1} I.~V.~Rakobolskaya, T.~M.~Roganova, L.~G.~Sveshnikovaa, Nucl. Phys. B~\textbf{122}, 353 (2003) 

\bibitem{cite:theory2} A.~Bhattacharya, R.~Enberg, M.~H.~Reno, I.~Sarcevica, A.~Stasto, JHEP \textbf{06}, 110 (2015) 

\bibitem{cite:cosmic1} T.~K.~Gaisser,  Astropart. Phys. \textbf{35}, 801 (2012) 

\bibitem{cite:cosmic2} T.~Stanev, T.~K.~Gaisser, S.~Tilav, Front. Phys. China \textbf{8}, 748 (2014)


\bibitem{cite:cascade-eq1} P.~Lipari, Astropart.Phys. \textbf{1}, 195 (1993)
 
\bibitem{cite:total-inelastic-p-air} P.~Abreu~\textit{et al.}~(Pierre Auger Observatory),  Phys. Rev. Lett. \textbf{109}, 062002 (2012)

\bibitem{cite:lhc1} R.~Aaij~\textit{et al.}~(LHCb Collaboration), JHEP \textbf{03}, 159 (2016) 

\bibitem{cite:lhc2} B.~Abelev~\textit{et al.}~(ALICE Collaboration), JHEP \textbf{07}, 191 (2012)

\bibitem{cite:lhc3} B.~Abelev~\textit{et al.}~(ALICE Collaboration), Phys. Lett. B \textbf{718}, 279 (2012)

\bibitem{cite:lhc4} B.~Abelev~\textit{et al.}~(ALICE Collaboration), JHEP \textbf{01}, 128 (2012)

\bibitem{cite:intrinsic-charm} S.~J.~Brodsky, A.~Kusina, F.~Lyonnet, I.~Schienbein, H.~Spiesberger, R.~Vogt, Adv. High Energy Phys. \textbf{2015}, 231547 (2015)

\bibitem{cite:diffractive-components} V.~D.~Barger, F.~Halzen, W.~Y.~Keung, Phys. Rev. D~\textbf{25}, 112 (1982)

\bibitem{cite:halzen-logan} F.~Halzen, L.~Wille, Phys. Rev. D~\textbf{94}, 014014 (2016)

\bibitem{cite:33} R.~Enberg, M.~H.~Reno, I.~Sarcevic, Phys. Rev. D~\textbf{78}, 043005 (2008) 

\bibitem{cite:spencer-gaisser} T.~Gaisser, S.~R.~Klein, Astro. Phys. \textbf{64}, 13 (2015)


\bibitem{cite:self-veto} T.~K.~Gaisser, K.~Jero, A.~Karle, J.~V.~Santen, Phys. Rev. \textbf{D 90}, 023009 (2014)

			
\bibitem{cite:nu_mu-6y} M.~G.~Aartsen~\textit{et al.}~(IceCube Collaboration), arXiv: \textbf{1607.08006} [astro-ph.HE], (2016)

\bibitem{cite:ers} R. Enberg, M. H. Reno, I. Sarcevic, Phys.Rev. D~\textbf{78}, 043005 (2008)

\bibitem{cite:sarkar} R.~Gauld, J.~Rojo, L.~Rottoli, S.~Sarkar, J.~Talbert, JHEP \textbf{02}, 130 (2016)


\bibitem{cite:pqcd_m_1} M.~V.~Garzelli, S.~Moch, G.~Sigl, JHEP \textbf{10}, 115 (2015)


\bibitem{cite:pqcd_m_2} A.~Bhattacharya, R.~Enberg, M.~H.~Reno, I.~Sarcevic, A.~Stasto, JHEP \textbf{06}, 110 (2015)

\bibitem{cite:muon}  M.~G.~Aartsen~\textit{et al.}~(IceCube Collaboration), arXiv: \textbf{1506.07981} [astro-ph.HE], (2016)

\bibitem{cite:muon-add} J.~I.~Illana, P.~Lipari, M.~Masip, D.~Meloni, Astropart. Phys. \textbf{34}, 663 (2011) 

\bibitem{cite:muon-add2} J.~I.~Illana, M.~Masip, D.~Meloni, JCAP \textbf{0909}, 008 (2009) 



\end{thebibliography}
%
%

\end{document}